\documentclass[sigconf]{acmart}
\usepackage{geometry}
\usepackage{booktabs}
\usepackage{multirow}
\usepackage{amsmath}
\usepackage{graphicx}
\usepackage[utf8]{inputenc}
\usepackage{enumitem}
\usepackage{xcolor}
\usepackage{fontawesome5}
\usepackage{soul}
\usepackage{xcolor}
\usepackage{colortbl}

\sethlcolor{yellow!30} 
\newcommand{\method}[0]{GLASS}
\newcommand{\best}[1]{\cellcolor{gray!15}\textbf{#1}}
\AtBeginDocument{%
  }

\copyrightyear{2026}
\acmYear{2026}
\setcopyright{cc}
\setcctype{by}
\acmConference[RecSys '26]{20th ACM Conference on Recommender Systems}{September 27--October 02, 2026}{Minneapolis, MN, USA}
\acmBooktitle{20th ACM Conference on Recommender Systems (RecSys '26), September 27--October 02, 2026, Minneapolis, MN, USA}
\acmDOI{10.1145/3773078.3831769}
\acmISBN{979-8-4007-2284-4/2026/09}

\begin{document}

\title{Coarse-to-Fine Long-term Interest Modeling for \\Generative Recommendation}

\author{Shiteng Cao}
\authornote{Both authors contributed equally to this research.}
\email{caost24@mails.tsinghua.edu.com}
\affiliation{%
  \institution{Shenzhen International Graduate School, Tsinghua University}
  \city{Shenzhen}
  \country{China}
}

\author{Junda She}
\authornotemark[1] 
\email{shejunda@bupt.edu.cn}
\affiliation{%
  \institution{Beijing University of Posts and Telecommunications}
  \city{Beijing}
  \country{China}
}

\author{Bin Zeng}
\authornotemark[1] 
\email{zengbin03@kuaishou.com}
\affiliation{%
\country{Kuaishou Inc., Beijing, China}
  \institution{Kuaishou Inc.}
  \city{Beijing}
  \country{China}
}

\author{Ji Liu}
\email{liuji06@kuaishou.com}
\affiliation{%
\country{Kuaishou Inc., Beijing, China}
  \institution{Kuaishou Inc.}
  \city{Beijing}
  \country{China}
}

\author{Chengcheng Guo}
\email{guochengcheng03@kuaishou.com}
\affiliation{%
\country{Kuaishou Inc., Beijing, China}
  \institution{Kuaishou Inc.}
  \city{Beijing}
  \country{China}
}

\author{Kuo Cai}
\email{caikuo@kuaishou.com}
\affiliation{%
\country{Kuaishou Inc., Beijing, China}
  \institution{Kuaishou Inc.}
  \city{Beijing}
  \country{China}
}

\author{Qiang Luo}
\authornote{Corresponding author}

\email{luoqiang@kuaishou.com}
\affiliation{%
\country{Kuaishou Inc., Beijing, China}
  \institution{Kuaishou Inc.}
  \city{Beijing}
  \country{China}
}

\author{Ruiming Tang}
\authornotemark[2] 
\email{tangruiming@kuaishou.com}
\affiliation{%
\country{Kuaishou Inc., Beijing, China}
  \institution{Kuaishou Inc.}
  \city{Beijing}
  \country{China}
}

\author{Han Li}
\email{lihan08@kuaishou.com}
\affiliation{%
\country{Kuaishou Inc., Beijing, China}
  \institution{Kuaishou Inc.}
  \city{Beijing}
  \country{China}
}

\author{Kun Gai}
\email{gai.kun@qq.com}
\affiliation{%
\country{Unaffiliated, Beijing, China}
  \institution{Unaffiliated}
  \city{Beijing}
  \country{China}
}
\author{Zhiheng Li}
\authornotemark[2] 
\email{zhhli@tsinghua.edu.cn}
\affiliation{%
  \institution{Shenzhen International Graduate School, Tsinghua University}
  \city{Shenzhen}
  \country{China}}

\author{Cheng Yang}
\authornotemark[2] 
\email{yangcheng@bupt.edu.cn}
\affiliation{%
  \institution{Beijing University of Posts and Telecommunications}
  \city{Beijing}
  \country{China}
}

\renewcommand{\shortauthors}{S. Cao et al.}

\begin{abstract}

Leveraging long-term user behavioral patterns is a key trajectory for enhancing the accuracy of modern recommender systems. 
Due to the quadratic complexity of attention mechanisms, existing GR models are typically confined to short interaction sequences.
While pioneer works have attempted to adapt Search-based Interest Models (SIM) to the generative context, they typically overlook the inherent hierarchical distinction of SIDs.
GR is fundamentally a coarse-to-fine generation task, where the initial SIDs (prefix) determine the broad semantic category and the subsequent SIDs (suffix) pinpoint the specific item.
Thus, our core insight is that the prefix and suffix of SIDs require distinct long-term signal injections.
To bridge this gap, we propose \textbf{\method}, a \textbf{G}enerative recommendation framework that integrates \textbf{L}ong-term user interests into the generative process vi\textbf{A} \textbf{S}IDTier and \textbf{S}emantic Search.
For the generation of SID prefix, we introduce SID-Tier, a module that maps long-term interactions into a unified interest vector to enhance the prediction of the initial SID token. 
SID-Tier leverages the compact nature of the semantic codebook to incorporate cross features between the user’s long-term history and candidate semantic codes. 
Furthermore, for the generation of SID suffix , we present semantic hard search, which utilizes generated coarse-grained semantic ID as dynamic keys to extract relevant historical behaviors, which are then fused via an adaptive gated fusion module to recalibrate the trajectory of subsequent fine-grained tokens. 
Extensive experiments on two large-scale real-world datasets, TAOBAO-MM and KuaiRec, demonstrate that \method\ outperforms state-of-the-art baselines. A two-week online A/B test on a short-video platform demonstrate that \method\ achieves significant gains in recommendation quality.
Our codes are publicly available at 
\href{https://anonymous.4open.science/r/GLASS-B56B/README.md}{this anonymous link} 
to facilitate further research in generative recommendation.
\end{abstract}





\maketitle

\section{INTRODUCTION}

Generative Recommendation (GR)~\cite{rajput2023Tiger, hou2025actionpiece, yi2025dualgr, li2025generative_survey_kuaishou} has garnered significant attention for its capability to leverage the powerful sequence modeling aptitudes of Transformers to capture complex user-item correlations directly.
Unlike traditional retrieval models, the GR paradigm re-imagines recommendation as an autoregressive generation process over hierarchical Semantic IDs (SIDs). 
Because of the complexity of the attention mechanism, the computational cost of GR models grows quadratically with the input sequence length. 
As a result, standard GR methods are often restricted to short interaction sequences for SID generation, failing to capture the rich signals buried within a user's long-term history.

In industrial scenarios, capturing long-term user behavioral patterns has emerged as a fundamental driver for enhancing prediction accuracy\cite{pi2020sim, yi2025dualgr, wu2025muse, guo2025miss}. 
Long-term behavior modeling is non-trivial because it inevitably introduces substantial noise and computational overhead.
To overcome these challenges, the industry has widely adopted a two-stage paradigm, exemplified by Search-based Interest Models (SIM)\cite{pi2020sim}. These methods typically utilize a General Search Unit (GSU) to search relevant sub-sequences before applying heavy attention mechanisms.

Although pioneer work\cite{yi2025dualgr} introduces the above search-based method into the generative recommendation paradigm, they overlook the hierarchical distinction of SIDs when modeling long-term behavior.
Specifically, generative recommendation paradigm represents items as hierarchical Semantic IDs (SIDs) and the recommendation process is inherently a coarse-to-fine generation task\cite{rajput2023Tiger,deng2025onerec}.
The generation of the first few SIDs determines the broad category of the target item, which necessitates modeling cluster-level preferences to lock in the correct semantic branch. 
Conversely, the subsequent SIDs are responsible for pinpointing the specific item within that cluster, requiring item-level preferences to distinguish fine-grained details against noise.
As a result, we argue that the generation of the first few SIDs (Prefix) and the subsequent SIDs (Suffix) should be treated differently.


To address these limitations, we propose \method, a \textbf{G}enerative recommendation framework designed to inject \textbf{L}ong-term interest vi\textbf{A} \textbf{S}IDTier and \textbf{S}emantic Search. 
Our approach leverages the properties of SIDs to enhance the generation of the initial and subsequent tokens respectively by injecting long-term user behavior.
For the initial token ($sid_1$), since no path has been established, we must proactively inject long-term interest. 
For subsequent tokens ($sid_{>1}$),  since $sid_1$ is determined, it functions as a semantic trigger that constrains the search space. This allows us to use the generated prefix as a key to search relevant information from the user’s history.

Specifically, for the first-level SID, we leverage the compact nature of the SID codebook. While traditional retrieval operates over massive, sparse item ID spaces, making complex cross features computationally prohibitive, the SIDs constrains the solution space. 
This efficiency makes it feasible to employ cross features during the retrieval stage.
Based on this insight, we introduce SID-Tier, a module designed to capture the intensity of a user’s interest across the semantic space.
By mapping long-term interactions into a unified interest vector, SID-Tier enhances the generation of $sid_1$ and ensures the autoregressive process starts on solid ground.

Then we conduct an analysis using the conditional rank progression metric to investigate the model’s trajectory during generation. Our findings reveal the error accumulation effect in current SID paradigms:after $sid_1$ is generated, the model’s focus on the ground-truth item often degrades rather than improves. Instead of the subsequent tokens ($sid_2, sid_3$) sharpening the search, the model frequently drifts away from the intended target as the sequence lengthens.
To bridge this gap, we utilize the generated first SID as keys to recalibrate the model's trajectory. Since $sid_1$ represents a coarse-grained semantic cluster, it acts as a latent target once generated. We implement semantic hard search: once $sid_1$ is determined, it acts as a key to retrieve relevant historical items. These items are fused via an adaptive gated fusion mechanism to refine the generation of fine-grained tokens.
A practical challenge in semantic hard search is the information scarcity bottleneck. If the generated $sid_1$ only maps to a limited number of interacted items, the semantic hard search module fails to provide sufficient information gain to guide the subsequent fine-grained generation. To address this, we propose two alternative strategies to ensure the retrieved context is informative: either (1) a relaxed matching mechanism that dynamically incorporates semantic neighbors, or (2) a structural resizing of the codebook to naturally increase the
density of bucket. Both approaches activate the potential of long-term history to recalibrate the model’s recommendation accuracy.

The contributions of this paper are summarized as follows:
\begin{itemize}[leftmargin=*]
    \item We propose \textbf{GLASS}, a novel framework designed to inject long-term behavior signals into the generative recomemender system,which outperforms other methods on two real-world datasets.
    \item We propose SID-Tier, a module that extracts global preferences to refine the generation of the initial SID.
    
    \item We present semantic hard search, leveraging generated SID as keys to search in long-term histories.
    
    \item Extensive experiments on real-world datasets demonstrate that \method\ consistently outperforms SOTA baselines. 
\end{itemize}

\section{RELATED WORK}

\subsection{Long Sequence User Interest Modeling}

To address the computational bottlenecks in capturing long-term user interests, the industry has widely adopted the two-stage paradigm pioneered by SIM.~\cite{pi2020sim}. SIM decomposes the modeling process into GSU and ESU. The GSU employs hard search or soft search to filter out subsequences relevant to the target item from the long history, which are  then modeled by the ESU. Subsequent works~\cite{chen2021eta,cao2022SDIM,chang2023twin,si2024twin-v2} have focused on optimizing efficiency and consistency. 
However, a critical distinction exists: the aforementioned advancements are predominantly designed for the ranking stage, where the target item is known. In the retrieval stage, the absence of a fixed target item during inference creates a query vacancy.
Recent studies have attempted to bridge this gap for retrieval. ULIM~\cite{meng2025ulim} proposes hierarchical clustering of long sequences and utilizes recent user interests as a query to model hierarchical interest points. LongRetriever~\cite{ren2025longretriever} designs the in context training paradigm, employing multi context vectors to simulate interactions between users and candidates. MISS ~\cite{guo2025miss} dynamically fuses long-sequence features to realize coarse-to-fine interest matching based on Tree-based Deep Models.
Despite these advancements, a critical research gap remains. These existing methods are primarily designed for traditional retrieval frameworks, which rely on calculating similarities between fixed embeddings. They do not translate directly to GR systems, where the challenge lies in maintaining autoregressive consistency over long sequences of Semantic IDs. Effectively integrating long-term history into the generative process,without succumbing to the quadratic complexity of attention,remains an unsolved challenge.

\subsection{Generative Recommender System}

Generative recommender systems \cite{li2025generative_survey_kuaishou, rajput2023Tiger, deng2025onerec} represent a paradigm shift where items are quantized into hierarchical Semantic IDs  (e.g., RQ-VAE or RQ-Kmeans). Within this framework, several domain-specific models have been proposed, such as OneRec for short-video recommendation \cite{deng2025onerec} , EGA~\cite{zheng2025ega} and GPR \cite{zhang2025gpr} for advertising systems, OneLoc \cite{wei2025oneloc} for local life services, and OneSearch, TBGRecall, FORGE for e-commerce systems \cite{chen2025onesearch, liang2025tbgrecall, fu2025forge}.

A core challenge in GR is to  model diverse user interaction patterns.
To capture a user’s immediate search intent, OneSearch~\cite{chen2025onesearch} explicitly injects the user's most recent search queries and clicked items into the input.  Similarly, OneRec~\cite{zhou2025onerec_v1} proposes a multi-scale behavior framework. The short-term pathway processes the most recent interactions , while the positive-feedback pathway tracks a sequence of  high-engagement interactions. To handle extended historical sequences,both frameworks employ a Q-Former architecture to compress these long-term behaviors into fixed-length latent vectors.
While these methods effectively condense lifelong trajectories, they operate at the \textit{interest-cluster} granularity rather than the \textit{item} granularity. While DualGR~\cite{yi2025dualgr} demonstrates that SID can effectively cluster items and uses the SID to act as the hard search key in long sequence modeling, it overlooks the inherent structural properties of SIDs. Consequently, the potential for SIDs to capture a more coherent, fine-grained representation of long-term evolution remains largely untapped. This leaves a critical gap in achieving a more effective way of modeling user's long behavior.
\begin{figure*}[t]
  \centering
  \includegraphics[width=\textwidth]{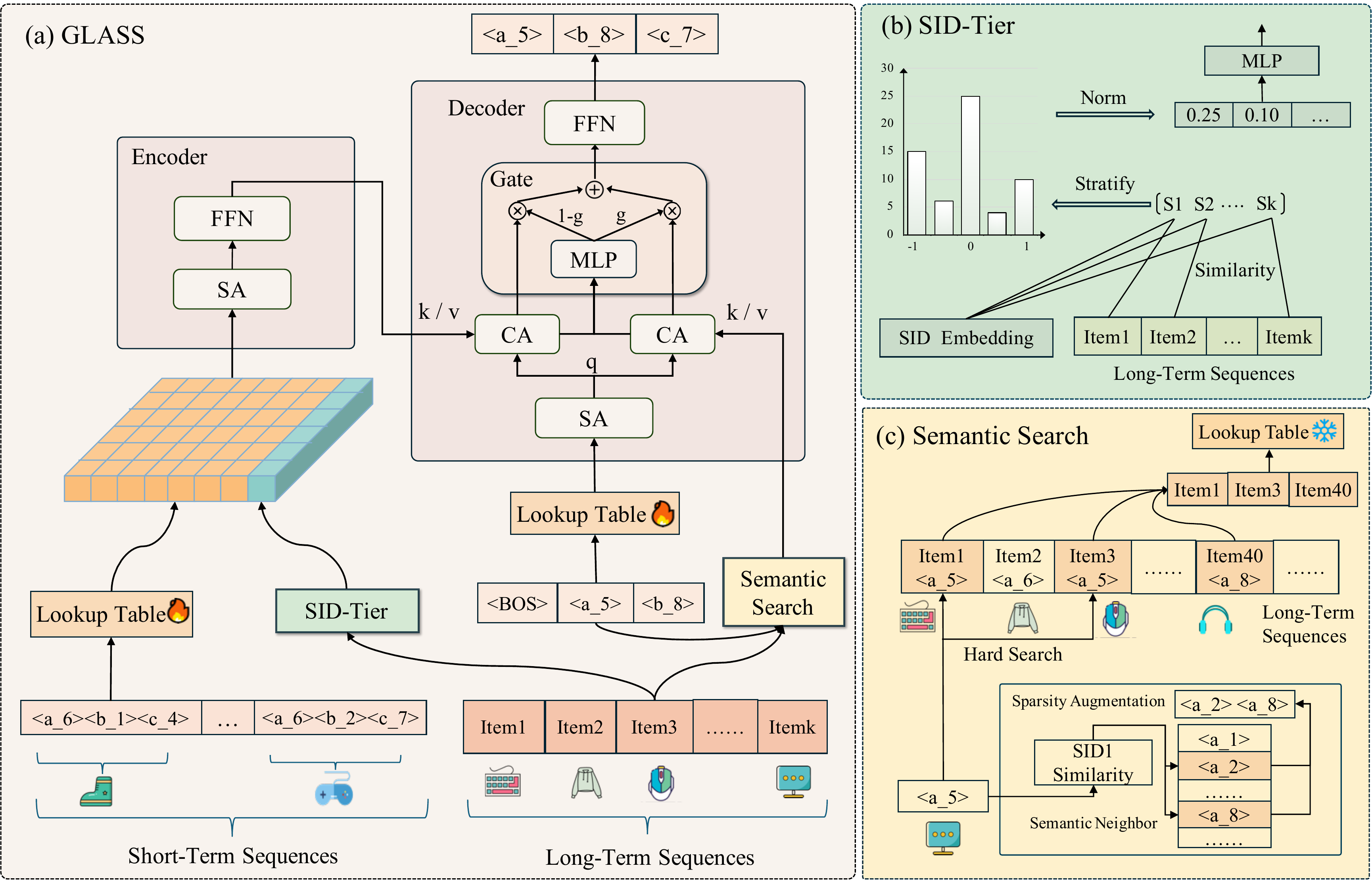} 
  \caption{Overview of the proposed \method\ framework. 
  (a) \textbf{GLASS} adopts the encoder-decoder architecture. The encoder processes short-term item sequences and SID-Tier to capture user interests. The decoder integrates encoder output and Semantic Search via dual cross attention.
  (b) \textbf{SID-Tier} aggregates long-term history into a unified interest vector to enhance the first codeword ($sid_1$) prediction. 
  (c) \textbf{Semantic Hard Search} performs  hard search based on the generated $sid_1$, followed by a gate to enhance the generation of fine-grained tokens ($sid_2, sid_3$). sparsity-aware augmentation ensures robust context by incorporating semantic neighbors or resizing the codebook.}
  \label{fig:overview}
\end{figure*}
\section{PRELIMINARY}

\textbf{Item feature.}
Let $\mathcal{V}$ be the set of items.
For each item $v \in \mathcal{V}$, we can transform it to its multi-modal embedding $\mathbf{h}_v$.
The multi-model embeddings are then quantized to generate a SID for each item.
Following Tiger\cite{rajput2023Tiger}, we use RQ-VAE for quantization and set the size of the $j$ level codebook $\mathcal{C}_j$ as $K_j$. 
Thus, we can transform each item $v$ to its SID $c(v)=(c_0(v),...,c_{m-1}(v))$, where $c_{j}(v)\in \mathcal{C}_j$ and the codebook number $m$ is $3$ in our method.
For the codeword in the $j$-th level $a\in \mathcal{C}_j$, it corresponds to a token in the Transformer block and the token embedding is $\mathbf{e}_a = Lookup(a) \in \mathbb{R}^d$ via the embedding lookup table.

\textbf{User interaction sequence.}
For each user, we construct the item sequence $H^{id}=(v_1,v_2,...,v_n)$ by chronologically sorting the interacted items, where $n$ is the sequence length.
Considering that the lifelong sequence can be excessively long, it is common to split the sequence into a long-term part $H^{id}_{long}=(v_1,...,v_l)$ and a short-term part $H^{id}_{short}=(v_{l+1},...,v_n)$, where $l$ is the length of the long-term sequence.
With the constructed SID, we convert all the id sequences into SID sequences, including $H=(c(v_1),c(v_2),...,c(v_n))$, $H_{long}$, and $H_{short}$.
Each codeword has its token embedding, and thus the embedding is $\mathbf{E}=[\mathbf{e}_{c(v_1)};...;\mathbf{e}_{c(v_n)}]\in \mathbb{R}^{nm\times d}$.
Similarly, we can also define $\mathbf{E}_{long}$ and $\mathbf{E}_{short}$.

Then, the recommendation task is to predict the next item $v_{n+1}$ according to the user's long-term and short-term behavior sequence:
\begin{equation}
    P(c(v_{n+1}) \mid H_{short}, H_{long}),
\end{equation}
where $c_{<j}(v_{n+1}) = (c_0, \dots, c_{j-1})$ represents the prefix codes generated within the current item step.

\section{METHODOLOGY}

As mentioned above, it is challenging to introduce long-term sequence into generative recommendation paradigm.
To solve this problem, we propose GLASS, a generative long-sequence modeling framework via SID-Tier and Semantic Search. 
To retrieve related items from the long-term sequence, we perform hard search with the first codeword.
Although the generation of follow-up codewords can be boosted by retrieved items, the generation of the first codeword lacks extra information.
Thus, we propose SID-Tier to capture the cross feature between the first codeword and the long-term sequence, which can be calculated in advance.
After that, we propose a gate module to better merge user's short-term and long-term interest.

The overall architecture of our proposed \method\ is illustrated in Figure \ref{fig:overview}. 
It consists of three primary components:  SID-Tier module to enhance the prediction of the initial token, a semantic hard search module for fine-grained trajectory recalibration, and a sparsity-aware augmentation strategy to  provide sufficient information gain to guide the subsequent fine-grained generation.

\subsection{SID-Tier}
\label{sec:sid_tier}

To effectively capture global user preferences over the first-level semantic space, we propose the SID-Tier module. 
Since the initial token generation lacks a preceding trajectory, SID-Tier aggregates long-term historical interactions into a context vector via a tier-aware quantization mechanism to serve as a "cold-start" guide.

\textbf{Prototype Alignment.} 
A critical challenge is the semantic misalignment between discrete codewords and continuous item embeddings. To bridge this gap, we assign a prototype embedding $\tilde{\mathbf{h}}_k \in \mathbb{R}^d$ to each codeword $k$ in the first-level codebook $\mathcal{C}_0$. This prototype is obtained by mean-pooling the embeddings of all items belonging to this cluster:
\begin{equation}
    \tilde{\mathbf{h}}_k = \frac{1}{|\mathcal{V}_k|} \sum_{v \in \mathcal{V}_k} \mathbf{h}_v, \quad \text{where } \mathcal{V}_k = \{v \in \mathcal{V} \mid c_0(v) = k\}.
\end{equation}
This ensures that the representation of each codeword is semantically centered within its corresponding item cluster.

\textbf{Interest Histogram Construction.}
We quantify the user's affinity towards each latent semantic category by computing the cosine similarity between the prototype $\tilde{\mathbf{h}}_k$ and every item in the user's long-term history $H_{long}^{id}$.
To capture the \textit{intensity distribution} of interest, we discretize the similarity range $[-1, 1]$ into $N$ intervals (tiers). For each codeword $k \in \mathcal{C}_0$, we construct a unified interest vector (histogram) $\mathbf{t}_k \in \mathbb{R}^N$:
\begin{equation}
    \mathbf{t}_k[j] = \sum_{v \in H_{long}^{id}} \mathbb{I}\left( \text{bin}(\text{Cos}(\tilde{\mathbf{h}}_k, \mathbf{h}_v)) = j \right),
\end{equation}
where $j \in \{1, \dots, N\}$ denotes the tier index, $\text{bin}(\cdot)$ maps a similarity score to its corresponding interval, and $\mathbb{I}(\cdot)$ is the indicator function. This vector $\mathbf{t}_k$ explicitly encodes the frequency of historical items falling into different affinity levels for a given semantic cluster.

\textbf{Global Injection.}
We concatenate the interest vectors of all first-level codewords to form a comprehensive heatmap of user interests. To align this high-dimensional feature with the token space, we flatten and project it using a Multi-Layer Perceptron (MLP):
\begin{align}
    \mathbf{x} &= \text{Concat}(\mathbf{t}_0, \mathbf{t}_1, \dots, \mathbf{t}_{K_0-1}) \in \mathbb{R}^{K_0 \cdot N}, \\
    \mathbf{e}_{tier} &= \text{MLP}(\mathbf{x}) \in \mathbb{R}^d,
\end{align}
where $K_0$ is the size of the first codebook. 
Finally, $\mathbf{e}_{tier}$ is appended to the user's short-term sequence embedding $\mathbf{E}_{short}$. The augmented sequence is then fed into the encoder:
\begin{equation}
    \mathbf{O}_{short}^{enc} = \text{Encoder}([\mathbf{E}_{short}; \mathbf{e}_{tier}]),
\end{equation}
where $[;]$ denotes the sequence concatenation. This allows the model to condition the generation of the initial token $sid_1$ on the global long-term preference profile.

\subsection{Decoding with Semantic Search}
In this section, we present the core architecture of semantic hard search, designed to dynamically integrate long-term historical preferences into the autoregressive generation process. 
Specifically, our approach leverages the hierarchical nature of semantic IDs to construct a semantic search mechanism, followed by an adaptive gated fusion strategy to balance short-term sequence and retrieved long-term sequence.

\textbf{Semantic search.} 
We adopt the hierarchical semantic IDs generated by RQ-VAE to represent each item $v$ as a discrete code sequence $c(v) = (c_{0}(v),...,c_{2}(v))$. This hierarchical structure naturally encodes semantic granularity.
Level-1 ($c_{0}(v)$) represents a coarse-grained semantic bucket. It inherently possesses clustering properties, grouping semantically similar items under the same ID.
Level-2/3 ($c_{1}(v),..., c_{2}(v)$) represents fine-grained semantics, progressively approaching a unique identifier of an item.

During the autoregressive decoding process, after the model generates the first codeword $\hat{c}_0(v_{n+1})$, we utilize this determined coarse intention as a semantic key to perform a hard search over the user's long-term history $H_{long}^{id}$. 
We construct a retrieved sequence $H_{ret}$ that contains only historical items that share the same first-level codeword as the generated one:
\begin{align}
    H_{ret} &= \{c(v_i) \mid v_i \in H_{long}^{id},  c_0(v_i) = \hat{c}_0(v_{n+1}) \}. 
\end{align}

This mechanism essentially provides the decoder with a Retrieval-Augmented Generation (RAG) context window. 
Unlike the standard RAG that might retrieve based on dense vector similarity, our approach filters the history based on the explicit semantic alignment of the coarse-grained intent ($\hat{c}_0(v_{n+1})$). This significantly reduces the interference of irrelevant noise by strictly limiting the context to historical behaviors that are semantically consistent with the current generation intent.

\textbf{Adaptive gate for fusion.}
While previous methods like DualGR employ a DBR for hard selection at the input level, we propose an adaptive gate in the decoder to fuse short-term and long-term interest. 
This strategy allows for a dynamic balance between short-term sequential patterns and long-term retrieved semantic signals.

The input of the decoder is $H_q^{dec}=[t_{bos};\hat{c}_0(v_{n+1});...;\hat{c}_{q-1}(v_{n+1})]$ in step $q$, where $t_{bos}=H_q^{dec}$ is the initial token.
With the above short-term and long-term interests, we employ two parallel cross attention modules to construct distinct views of the user's preference:
\begin{align}
& \mathbf{E}_{ret} = Lookup(\mathbf{H}_{ret}) \\
& \mathbf{X} = SelfAttn(Lookup(\mathbf{H}_q^{dec})), \\
& \mathbf{Z}_{short} = CrossAttn_{\theta}(\mathbf{X}, \mathbf{O}_{short}^{enc}, \mathbf{O}^{enc}_{short}), \\
& \mathbf{Z}_{ret} = CrossAttn_{\phi}(\mathbf{X}, \mathbf{E}_{ret}, \mathbf{E}_{ret}),
\end{align}
where $SelfAttn(\cdot)$ is the self attention module with residual connections and normalization\cite{vaswani2017attention}, $CrossAttn_\theta(\cdot, \cdot, \cdot)$ is the cross attention module with the first parameter as query and the other parameters as key and value, $Lookup$ is the embedding lookup table and $\mathbf{O}_{short}^{dec}$ is the output of the encoder.
The short-term View (${H}_{short}$) captures sequential patterns and immediate intent from recent interactions.
The retrieved View (${H}_{ret}$) aggregates long-term preference information related to the current generated semantic.

To effectively integrate these views, we introduce a gate that acts as a self-verification module. It evaluates the relevance of the retrieved information relative to the current decoding state. We compute a scalar gate value $g \in [0, 1]^{nm \times d}$ by projecting the concatenated representations:

\begin{equation}
\mathbf{g} = \sigma([\mathbf{Z}_{short} \ ; \ \mathbf{Z}_{ret}]\mathbf{W}_g )    
\end{equation}

where $[ ; ]$ denotes the concatenation operation, and $\mathbf{W}_g \in \mathbb{R}^{2d \times d}$ is the gate weight.
The final context vector $\mathbf{Z}_{context}$ is obtained by weighted summation:

\begin{equation}
\mathbf{Z}_{context} = (\textbf{1} - \mathbf{g}) \odot \mathbf{Z}_{short} + \mathbf{g} \odot \mathbf{Z}_{ret},    
\end{equation}
where $\mathbf{1}$ is a $d$-dimension vector of ones.
The gate weight vector $\mathbf{g}$ controls the contribution of retrieval signals: a 'large' $\mathbf{g}$ emphasizes the retrieved long-term semantics $\mathbf{Z}_{ret}$, while a 'small' $\mathbf{g}$ suppresses retrieval and relies on the short-term behavior $\mathbf{H}_{short}$.

The gated context $\mathbf{Z}_{context}$ is then processed by a position-wise Feed-Forward Network (FFN) accompanied by residual connections and layer normalization, producing the final hidden state of the current decoder layer.
By stacking $L$ decoder layers, we obtain the final hidden representation $\mathbf{h}_{q}^{dec}$ for the $q$-th step.

To generate the next token, we project $\mathbf{h}_{q}^{dec}$ into the codebook space via a linear head and compute the probability distribution $p(\hat{c}_{j}(v_{i}) | \hat{c}_{<j}(v_{i}), c(v_{<i}))$ over the codebook $\mathcal{C}_j$.
Finally, the model is trained by minimizing the negative log-likelihood loss over the hierarchical codeword sequence:
\begin{equation}
    \mathcal{L} = - \sum_{i=1}^{n} \sum_{j=0}^{m-1} \log p(\hat{c}_{j}(v_{i}) | \hat{c}_{<j}(v_{i}), c(v_{<i})),
\end{equation}
where $m=3$ denotes the number of quantization levels (i.e., SID depth).


\subsection{Sparsity-Aware Augmentation}
\label{sec:sparsity_aware}

Semantic Hard Search based on SID1 can efficiently extract relevant interest sequence based on semantic information, however, directly using this method may lead to a sparsity issue for the sub-sequences associated with the first SIDs. 
Consider a long-sequence recommendation setting where the user sequence length is $L=1000$ and the first-level codebook size is $|\mathcal{C}_1|=128$. Theoretically, each SID retrieves an average of only $1000/128 \approx 7.8$ historical items. Such a limited context window provides negligible Information gain and is highly susceptible to noise. Even in industrial deployments, where history lengths increase, the codebook size typically scales proportionally, rendering the insufficient retrieval length a persistent bottleneck. To alleviate this and enhance the robustness of the retrieval context, we propose two complementary strategies: \textit{ Semantic Neighbor Augmentation} and \textit{Codebook Resizing}.

\subsubsection{ Semantic Neighbor Augmentation}
This strategy introduces a Relaxed Matching Mechanism that leverages the semantic topology of the codebook to augment sparse contexts.
Based on the code embeddings learned via Residual Quantization, we compute the pairwise cosine similarity between all codes in the first layer. For each code $c_i$, we construct a static dictionary by recording its top-$k$ nearest semantic neighbors .
During the generation process, when $SID_1$ is produced to retrieve historical items, we employ a dynamic length verification mechanism. Let $R(SID_1)$ denote the set of retrieved items. If the sequence length $|R(SID_1)|$ falls below a predefined threshold $\tau$, the system triggers the Neighbor Augmentation protocol. We activate the pre-computed nearest neighbors of $SID_1$ and merge their associated historical items into the current retrieval set. This approach enhances the sequence length and information density while maintaining semantic consistency.

\subsubsection{Codebook Resizing}
Our second strategy is to optimize structure of the semantic code to increase the density of retrieved sequence .
We constrain the size of the first-level codebook $|\mathcal{C}_1|$ during training, forcing the model to aggregate a broader set of items into each SID bucket. While conventional wisdom suggests that altering codebook hyperparameters yields diminishing returns—as it typically only re-allocates prediction difficulty between layers—this adjustment catalyzes a qualitative paradigm shift within our hard semantic search framework. 
By compacting the first-level codebook, we drastically increase the average item count per SID bucket, ensuring the retrieved context possesses the requisite information volume for robust augmentation.By decoupling layer responsibilities, assigning density to the primary layer and precision to subsequent layers, we effectively allievate the sparsity bottleneck. 

\section{EXPERIMENTS}

To comprehensively evaluate the effectiveness of \method, our experiments are designed to research the following research questions
\begin{itemize}[leftmargin=*]
    \item \textbf{RQ1:} How does \method\ perform compared to state-of-the-art baselines?
    \item \textbf{RQ2:} What is the contribution of the core components to the overall performance?
    \item \textbf{RQ3:} How does the learnable gating mechanism balance noise and signal in retrieved sequences?
    \item \textbf{RQ4:} Why is Semantic Hard Search effective?
    \item \textbf{RQ5:} How does \method\ perform in Industrial dataset?
\end{itemize}

\subsection{Experimental Settings}

\subsubsection{Dataset Description.} 

We conduct experiments on two real-world datasets: \textbf{TAOBAO-MM}~\cite{liu2024multimodal-survey-kdd} and \textbf{KuaiRec}~\cite{Gao2022KuaiRec}. 
Since both datasets are originally designed for ranking tasks, we adapt them for long-sequence retrieval by treating only positive interactions as input sequences. 
Specifically, for TAOBAO-MM, we retain user-item interactions with \textit{clicks}; for KuaiRec, we select samples where the watch ratio (defined as $\text{play time} / \text{video duration}$) exceeds $100\%$.

Given the large item vocabulary in TAOBAO-MM, we define a target domain by first uniformly sampling 10,000 items and then filtering to retain only those that appear at least five times in the training set. This ensures sufficient statistical support for learning stable item representations.
Following the configuration in MUSE~\cite{liu2024multimodal-survey-kdd}, we partition each user’s interaction history: the most recent 50 interactions are treated as the short-term sequence, and all earlier interactions are grouped into the long-term history.

\begin{table}[htbp]
\centering
\caption{Statistics of the TAOBAO-MM and KuaiRec datasets after preprocessing.}
\label{tab:dataset_stats}
\begin{tabular}{lcc}
\toprule
\textbf{Statistic} & \textbf{TAOBAO-MM} & \textbf{KuaiRec} \\
\midrule
\# Users & 189,423 & 3790 \\
\# Items & 10,000 & 10,728 \\
\# Training samples & 265,301 & 30,320 \\
\# Validation samples & 31,205 & 3,790\\
\# Test samples & 31,206 & 3,790 \\
Avg historical sequence length & 987 & 884 \\
\bottomrule
\end{tabular}
\end{table}

\subsubsection{ Implementation Details.}
The hyperparameter configurations for experiments on public datasets are described as follows.  
Shared configurations for both SID-based and ID-based models include: embedding table dimension of 96, dropout rate of 0.1, AdamW optimizer with a learning rate of \(1 \times 10^{-4}\), short sequence length of 50, early stopping patience of 50, 8 attention heads, and key/value dimension of 32.
Task-specific configurations: for the SID-based model, the input representation uses semantic tokens with a codebook structure of \([64, 128, 128]\), a SID-tier MLP dimension of 256, and 4 encoder/decoder blocks; for the ID-based model, the input representation uses atomic item IDs, 20 negative samples, and 2 encoder/decoder blocks.



\subsubsection{Evaluation Metrics.}
To evaluate our model's performance, we utilize two widely-used ranking metrics: Recall@K, NDCG@K~\cite{ndcg}.

\subsubsection{Baselines.}
In our work, we evaluate the performance of our model with two groups of baselines to examine its effectiveness.

\noindent\textit{(1) ID based retrieval baselines}
\begin{itemize}[leftmargin=*]
    \item \textbf{Caser}~\cite{tang2018caser} introduces convolutional neural network  to capture Markovian patterns by applying  convolutions. 
    \item \textbf{HGN}~\cite{ma2019HGN} proposes a Hierarchical Gating Network that employs a gating mechanism to weight temporal signals.
    \item \textbf{GRU4Rec}~\cite{hidasi2016GRU4Rec}leverages a customized Gated Recurrent Unit for modeling sequential interactions. 
    \item \textbf{SASRec}~\cite{kang2018sasrec} adopts a causal self-attention Transformer to model user sequences.
    \item \textbf{Bert4Rec}~\cite{sun2019bert4rec} employs a bi-directional Transformer for next-item prediction.
    \item \textbf{S$^3$-Rec}~\cite{Zhou_2020_S3Rec}  pre-trains a bi-directional Transformer on self supervision tasks to improve sequential recommendation.
    \item \textbf{HSTU}~\cite{zhai2024hstu} replaces standard Transformer blocks with pointwise aggregation and spatial transformation layers.
\end{itemize}

\noindent\textit{(2) SID based retrieval baselines}
\begin{itemize}[leftmargin=*]
    \item \textbf{Tiger}~\cite{rajput2023Tiger} proposes a generative retrieval framework that represents items via Semantic IDs and uses Transformer to predict the Semantic ID of the next item.
 
    \item \textbf{DualGR}~\cite{yi2025dualgr} employs average pooling of history item embeddings and adopts a routing mechanism that guides the decoding to be conditioned on either the long-term or short-term branch. 
\end{itemize}

\subsection{Overall Performance Comparison (RQ1)}

\begin{table*}[htbp]
\centering
\caption{Overall performance comparison on Taobao-MM and KuaiRec datasets. The best results are highlighted in \textbf{bold}, and the second-best are \underline{underlined}. The background color indicates our proposed model.}
\label{tab:performance_new}
\renewcommand{\arraystretch}{0.95} 
\setlength{\tabcolsep}{5.5pt} 

\definecolor{Gray}{gray}{0.92}

\resizebox{0.95\textwidth}{!}{%
\begin{tabular}{cclccccccccc}
\toprule
\multirow{2}{*}{Dataset} & \multirow{2}{*}{Type} & \multirow{2}{*}{Model} & \multicolumn{5}{c}{Recall (R@K)} & \multicolumn{4}{c}{NDCG (N@K)} \\
\cmidrule(lr){4-8} \cmidrule(lr){9-12}
& & & R@1 & R@3 & R@5 & R@10 & R@20 & N@3 & N@5 & N@10 & N@20 \\
\midrule

\multirow{12}{*}{\rotatebox{90}{Taobao-MM}} 
& \multirow{7}{*}{ID} 
  & Caser    & 0.0191 & 0.0253 & 0.0293 & 0.0394 & 0.0611 & 0.0228 & 0.0243 & 0.0275 & 0.0330 \\
& & HGN      & 0.0179 & 0.0287 & 0.0340 & 0.0444 & 0.0575 & 0.0242 & 0.0264 & 0.0298 & 0.0329 \\
& & GRU4Rec  & 0.0196 & 0.0304 & 0.0397 & 0.0489 & 0.0587 & 0.0242 & 0.0296 & 0.0326 & 0.0350 \\
& & SASRec   & 0.0221 & 0.0386 & 0.0387 & 0.0524 & 0.0624 & 0.0321 & 0.0322 & 0.0365 & 0.0454 \\
& & Bert4Rec & 0.0267 & 0.0376 & 0.0475 & 0.0542 & 0.0731 & 0.0337 & 0.0374 & 0.0395 & 0.0443 \\
& & S$^3$-Rec& 0.0289 & 0.0347 & 0.0375 & 0.0552 & 0.0762 & 0.0321 & 0.0333 & 0.0453 & 0.0590 \\
&   & HSTU     & 0.0274 & 0.0452 & 0.0512 & 0.0784 & 0.1023 & 0.0384 & 0.0412 & 0.0495 & 0.0592 \\
\cmidrule{2-12}
& \multirow{3}{*}{SID} 

& Tiger    & 0.0280 & 0.0571 & 0.0756 & 0.1076 & \underline{0.1498} & \underline{0.0448} & 0.0523 & 0.0627 & 0.0733 \\
& & DualGR   & \underline{0.0306} & \underline{0.0592} & \underline{0.0781} & \underline{0.1126} & 0.1487 & 0.0438 & \underline{0.0578} & \underline{0.0653} & \underline{0.0738} \\
& & \best{GLASS} & \best{0.0372} & \best{0.0720} & \best{0.0934} & \best{0.1284} & \best{0.1723} & \best{0.0582} & \best{0.0669} & \best{0.0782} & \best{0.0893} \\
\cmidrule{2-12}
& \multicolumn{2}{c}{Impr.}   & \textbf{21.57\%} & \textbf{21.62\%} & \textbf{19.59\%} & \textbf{14.03\%} & \textbf{15.02\%} & \textbf{29.91\%} & \textbf{15.74\%} & \textbf{19.75\%} & \textbf{21.00\%} \\
\midrule \midrule

\multirow{12}{*}{\rotatebox{90}{KuaiRec}} 
& \multirow{7}{*}{ID} 
  & Caser    & 0.0185 & 0.0556 & 0.0603 & 0.0918 & 0.1254 & 0.0373 & 0.0392 & 0.0493 & 0.0578 \\
& & HGN      & 0.0262 & 0.0570 & 0.0791 & 0.1110 & 0.1582 & 0.0436 & 0.0525 & 0.0629 & 0.0747 \\
& & GRU4Rec  & 0.0348 & 0.0469 & 0.0582 & 0.0887 & 0.1263 & 0.0428 & 0.0475 & 0.0572 & 0.0666 \\
& & SASRec   & 0.0352 & 0.0638 & 0.0857 & 0.1180 & 0.1040 & 0.0498 & 0.0587 & 0.0703 & 0.0826 \\
& & Bert4Rec & 0.0404 & 0.0588 & 0.0698 & 0.0902 & 0.1244 & 0.0507 & 0.0553 & 0.0617 & 0.0704 \\
& & S$^3$-Rec& 0.0385 & 0.0625 & 0.0780 & 0.1125 & 0.1450 & 0.0535 & 0.0605 & 0.0710 & 0.0795 \\
&& HSTU     & 0.0334 & 0.0645 & 0.0845 & 0.1183 & \underline{0.1689} & 0.0513 & 0.0595 & 0.0704 & 0.0830 \\
\cmidrule{2-12}
& \multirow{3}{*}{SID} 

& Tiger    & 0.0406 & 0.0700 & 0.0859 & 0.1202 & 0.1530 & 0.0574 & 0.0641 & 0.0751 & 0.0834 \\
& & DualGR   & \underline{0.0445} & \underline{0.0745} & \underline{0.0932} & \underline{0.1292} & 0.1678 & \underline{0.0632} & \underline{0.0718} & \underline{0.0821} & \underline{0.0908} \\
& & \best{GLASS} & \best{0.0467} & \best{0.0778} & \best{0.0971} & \best{0.1345} & \best{0.1742} & \best{0.0659} & \best{0.0749} & \best{0.0855} & \best{0.0946} \\
\cmidrule{2-12}
& \multicolumn{2}{c}{Impr.}   & \textbf{4.94\%} & \textbf{4.43\%} & \textbf{4.18\%} & \textbf{4.10\%} & \textbf{3.81\%} & \textbf{4.27\%} & \textbf{4.32\%} & \textbf{4.14\%} & \textbf{4.19\%} \\
\bottomrule
\end{tabular}%
}
\end{table*}

Table~\ref{tab:performance_new} reports the overall performance of our proposed method and the baselines on the Taobao-MM and KuaiRec datasets. We observe that \method\ consistently outperforms all baseline models across all evaluation metrics on both datasets. Specifically, compared with the strongest baseline, our model achieves a significant relative improvement of up to 21.57\% and 29.91\% in terms of H@1 and N@3 on the Taobao-MM dataset. 

We attribute the superiority of \method\ to several key factors. 
First, compared to traditional sequential recommendation methods, \textsf{Ours} effectively leverages long-term user behavior sequences to capture more comprehensive user interests. 
Although the recent generative model \textsf{DualGR} also incorporates long-sequence information, it exhibits two major limitations that \textsf{\method} successfully addresses: 
(1) Instead of relying on simple average pooling to generate user representations, which fails to capture fine-grained interaction features between the user and the candidate items, our \textsf{SID-Tier} mechanism provides richer information by explicitly modeling target-aware interactions. 
(2) In the generation process of hierarchical SIDs, traditional method employs only a single cross attention module per layer, which limits its ability to organically fuse short-term and long-term behavioral signals. Our method adopts a more sophisticated fusion strategy, ensuring that both transient intents and persistent preferences are well-integrated.

Furthermore, we observe that the performance gain is more pronounced on the \textsf{Taobao-MM} dataset than on \textsf{KuaiRec}. This discrepancy stems from the difference in multimodal data quality. In the \textsf{Taobao-MM} dataset, we utilize high-quality multimodal features learned via supervised contrastive learning, which offers superior semantic alignment. In contrast, the multimodal information in \textsf{KuaiRec} is derived from simple text-based encoding, leading to relatively lower representational quality and thus more limited performance improvements. 

\subsection{Ablation Study (RQ2)}

\subsubsection{Synergy of SID-Tier and SHS}
\label{sec:ablation_study_synergy}

\begin{figure}[htbp]
  \centering
  \includegraphics[width=0.9\linewidth]{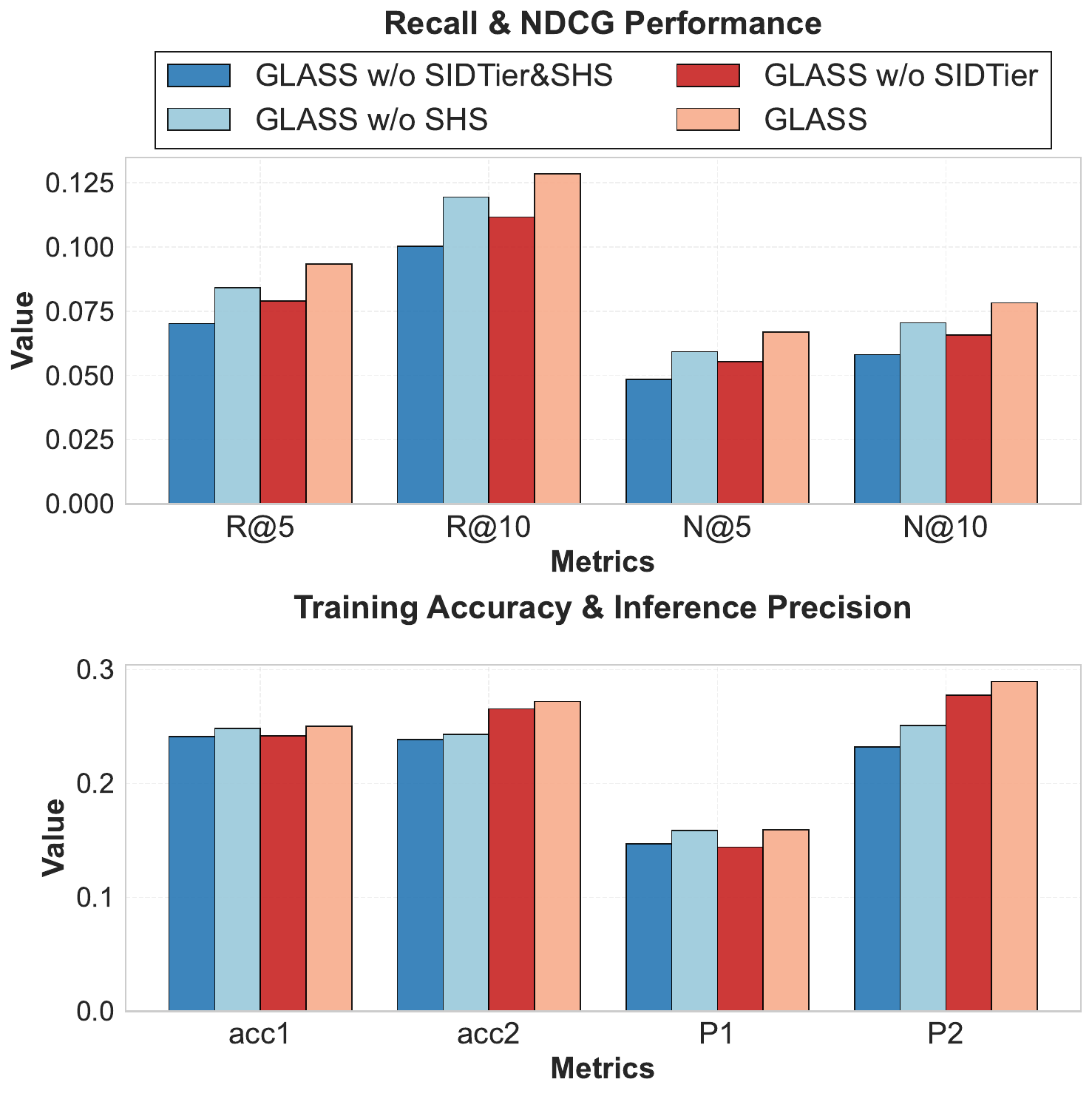}
  \caption{Performance comparison of SID-Tier and SHS modules. The legends represent: Vanilla (Blue), +SID-Tier (Light Blue), +SHS (Red), and GLASS Full (Orange).}
  \label{fig:performance_metrics}
\end{figure}

To investigate the specific locations and stages where different modules contribute to performance improvements , we analyze the model across two dimensions: Training Accuracy($acc_i$), measured under teacher-forcing to represent the model’s fit at the $i$-th hierarchical level, and Inference Precision ($P_i$), which measures the conditional probability that the $i$-th code is correct given that all preceding prefixes were accurately predicted during top-1 beam search.
As illustrated in Figure \ref{fig:performance_metrics}, the integration of the SID-Tier module yields substantial performance gains over the vanilla model. For instance, Recall@5 increases from 0.0702 to 0.0842. This improvement is fundamentally rooted in the prefix-dependent nature of the Semantic ID paradigm. By employing SID-Tier, $acc_1$ rises from 0.2410 to 0.2485, which directly translates to an increase in $P_1$ from 0.1470 to 0.1587.
The Semantic Hard Search (SHS) module provides a complementary boost. As shown in the training accuracy metrics in Figure \ref{fig:performance_metrics}, GLASS, which integrates both SID-Tier and SHS, promotes $acc_2$ and $P_2$ significantly. The synergy between SID-Tier and SHS allows the model to maintain high fidelity across the entire sequence, resulting in the highest observed values for Recall@10 and NDCG@10.

\subsubsection{Sparsity-Aware Augmentation Method}
\label{sec:ablation_study_sparsity}

\begin{figure}[htbp]
  \centering
  \includegraphics[width=0.9\linewidth]{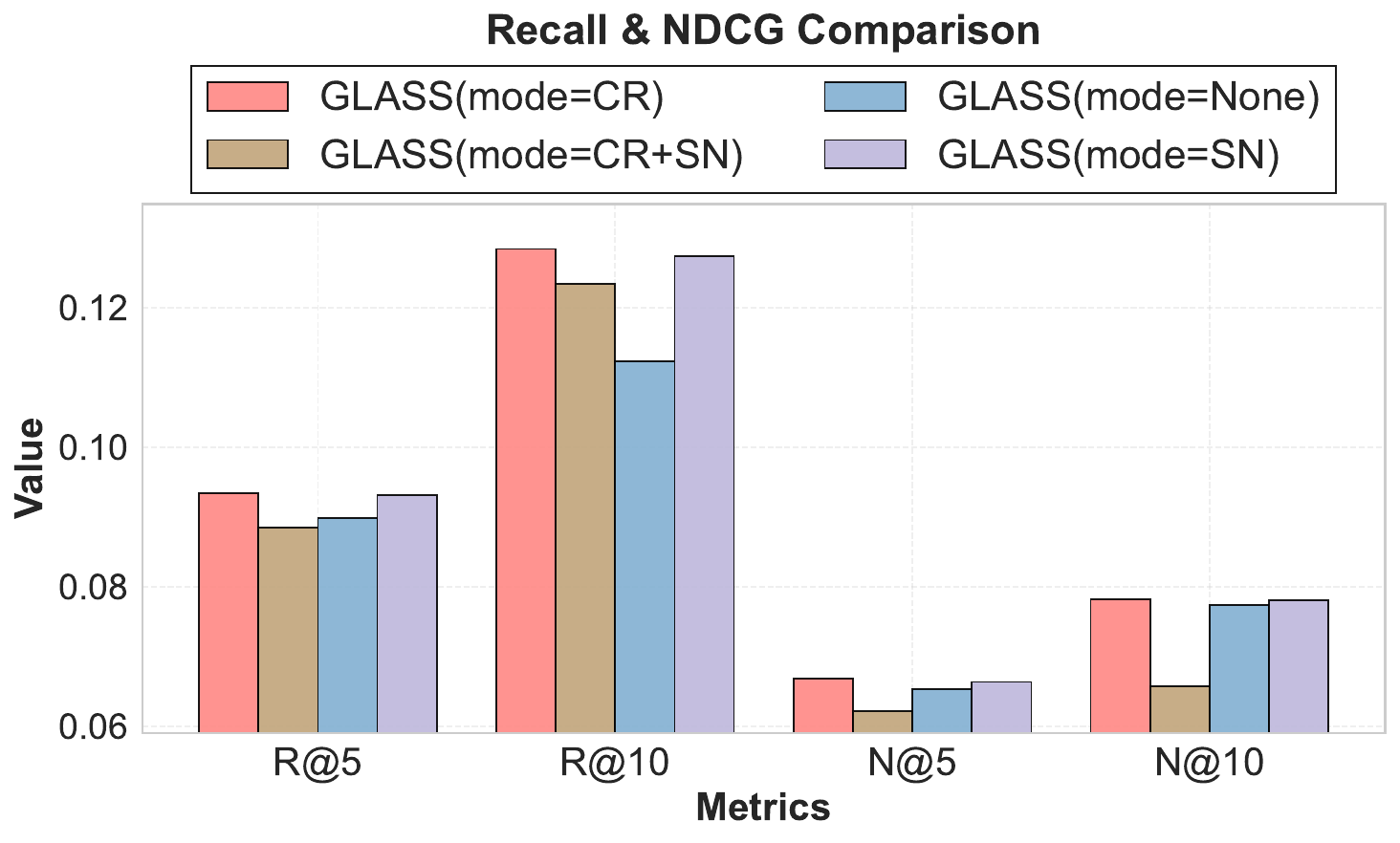}
  \caption{Ablation study on Sparsity Augmentation across different modes (None, CR, SN, and CR+SN).}
  \label{fig:sparsity_ablation}
\end{figure}

The impact of Codebook Resizing (CR) and Semantic Neighbor Augmentation (SN) is summarized in Figure \ref{fig:sparsity_ablation}. We observe that CR significantly enhances the model's ranking capability, with the GLASS(mode=CR) configuration achieving the peak Recall@10.
Regarding the SN strategy, the results in Figure \ref{fig:sparsity_ablation} exhibit divergent behavior. SN demonstrates its efficacy in the setting without codebook resizing, which confirms that SN can alleviate the sparsity bottleneck of SHS by augmenting the retrieval context with semantic neighbors.
Conversely, in settings with codebook resizing, SN shows a marginal degradation in performance compared to the mode without SN. This is attributed to the noise introduced by semantic neighbors; when codebooks represent broad interest clusters, softening the search space likely blurs the boundaries between distinct user interests.


\subsection{Impact of the Gating Mechanism(RQ3)}

To investigate the effectiveness of the proposed learnable gating mechanism compared to a static weighting approach, we conducted a comparative analysis. While a fixed weight (e.g., $0.5$) can offer improvements when the first-level codebook size is small, it leads to a performance degradation as the codebook granularity increases. This is primarily due to the high variance inherent in the retreived sequences. During inference, if the retrieved sequence length is too short, the signal-to-noise ratio decreases significantly. Under a fixed-weight scheme, these noisy signals directly distort the cumulative probabilities during beam search for subsequent levels, ultimately harming the final ranking performance.

\begin{figure}[h]
    \centering
    \includegraphics[width=\linewidth]{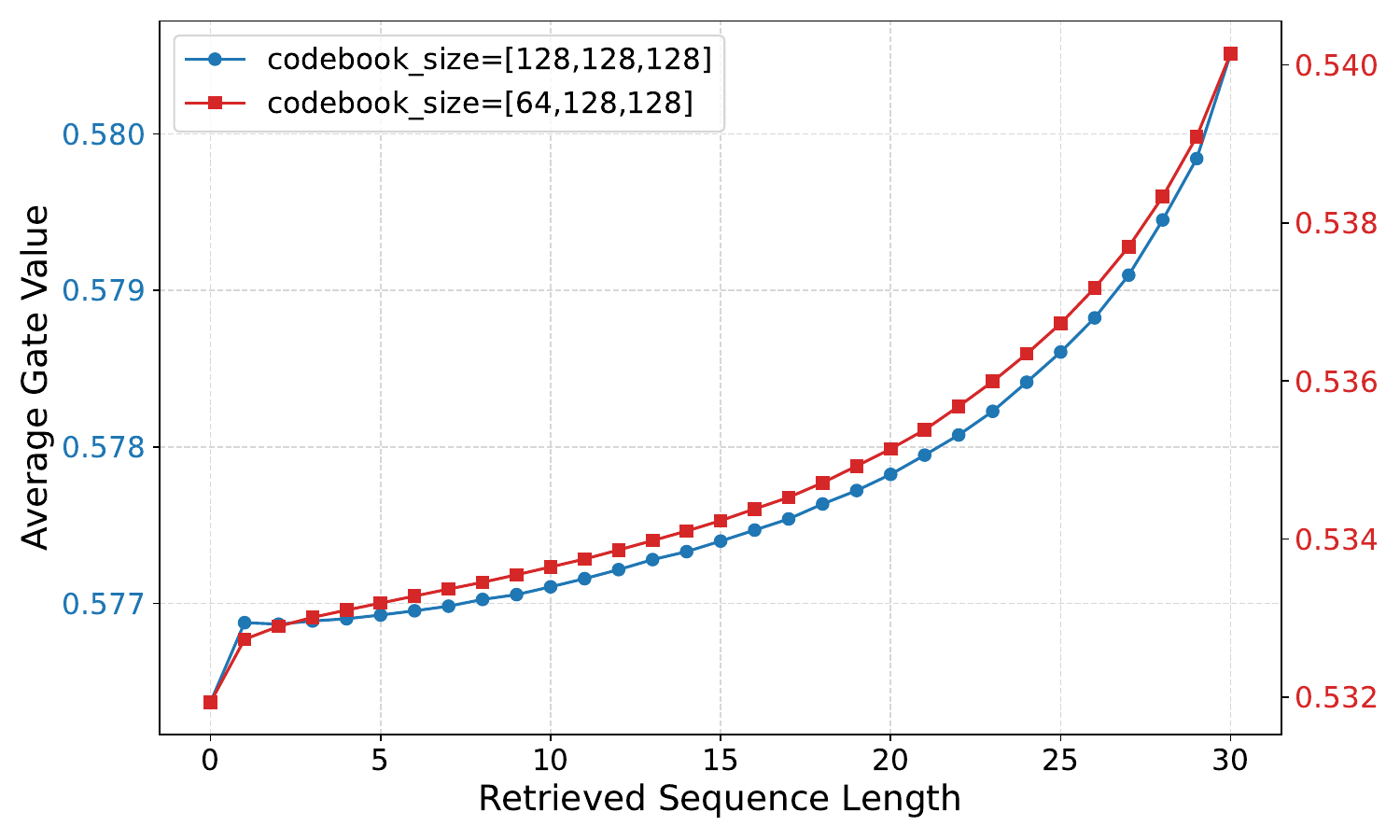}
    \caption{Average Gate Value vs. Retrieved Sequence Length during Inference. The curves represent two codebook configurations: $[128, 128, 128]$ (left) and $[64, 128, 128]$ (right).}
    \label{fig:gate_analysis}
\end{figure}

Furthermore, we analyze the behavior of the gate value relative to the retrieved sequence length using the first-level SID. As illustrated in Figure~\ref{fig:gate_analysis}, both codebook configurations exhibit a consistent upward trend: as the length of the retrieved sequence increases, the average gate value rises. This suggests that the model learns to assign higher importance to retrieved sequences as they generally become more informative. These results validate the effectiveness of our gating mechanism in dynamically balancing short-term and long-term interests to optimize the prediction of SID2 and SID3.

\subsection{Theoretical Justification for SHS(RQ4)}

To understand Semantic Hard Search, we analyze how ranking precision evolves during decoding. We use the CRP metric, which measures the expected rank of the ground-truth prefix within the beam at depth $d$, given it survived previous steps. 
To quantify the model's discriminative precision at varying semantic depths, we propose the \textbf{Conditional Rank Progression} (CRP). 
Conditioning on samples whose ground-truth prefix survives in the beam up to depth $d-1$, 
$\text{CRP}_d$ measures the expected rank of the correct $d$-th prefix among the top-$K$ candidates at step $d$:
\[
\text{CRP}_d = \frac{1}{|\mathcal{E}_d|} \sum_{n \in \mathcal{E}_d} R_d^{(n)},\quad 
\mathcal{E}_d = \left\{ n \mid \prod_{i=1}^{d-1} H_i^{(n)} = 1 \right\},\quad 
\mathcal{E}_1 = \mathcal{N}.
\]
A $\text{CRP}_d$ close to $1$ indicates high confidence, whereas a value near $K$ signals vulnerability to subsequent retrieval failure.
Ideally, hierarchical SIDs enable deep thinking, but we observe a \emph{rank degradation} phenomenon~\cite{guo2026prm} in vanilla generative retrieval. With beam size $K=20$, the baseline GR model increases CRP by $1.25$ at the second layer, while SHS limits it to $0.98$ (a $22\%$ improvement). At the third layer, SHS reduces degradation by $31\%$ (from $0.19$ to $0.13$).
The root cause lies in the probability chain of beam search. Given user context, the probability of semantic ID path is:
\begin{equation}
P(\text{path}_i \mid \text{context}) = P(sid_1 \mid \text{context}) \cdot \prod_{t=2}^{3} P(sid_t \mid sid_1,\dots,sid_{t-1}, \mathcal{C})
\end{equation}
In the baseline, $\mathcal{C}$ is static (only the initial context), failing to widen the probability gap between ground truth and distractors at finer granularities. SHS dynamically expands $\mathcal{C}$: for $n>1$, $\mathcal{C} = \{\text{context}, \text{context}'\}$, where $\text{context}'$ is a set of relevant items retrieved from the user's long-term history using the predicted $sid_1$ as a query. The initial context provides broad behavioral signals; $\text{context}'$ supplies fine-grained personal preferences, calibrating transition probabilities and mitigating rank degradation.

\subsection{Results on Industrial-scale Dataset (RQ5)}
To validate the practical value of our method in an industrial setting, we conducted an online A/B test on a short-video recommendation platform for two weeks. We allocated 3\%(approximately 12 million) of total users to the experimental group and another 3\% to the control group. Results show that \method\ significantly improves user engagement. At the 95\% confidence level, the experimental group achieved a +0.056\% increase in total app usage time and a +0.052\% increase in app usage time per user. Moreover, recall improved by 3 percentage points (absolute) compared to the baseline. These results strongly demonstrate the practical value of our method in industrial recommendation scenarios.

\section{CONCLUSION}

In this paper, we present \method, a novel generative recommendation framework designed to effectively harness long-term user interaction sequences within the generative recommender system. Our approach addresses the fundamental challenge of information under-utilization in existing GR systems when dealing with extensive behavioral histories.
We introduce SID-Tier, a module that facilitates feature interaction between a user's long history and the semantic codebook. By capturing global semantic preferences, SID-Tier provides a robust global context that enhances the prediction of the critical initial token.
Furthermore, we propose semantic hard search. By utilizing the generated SID as a search key, the model recalibrates the generation trajectory of subsequent fine-grained tokens. To overcome the information scarcity bottleneck in semantic search, we implement sparsity-aware augmentation, ensuring an informative context for fine-grained generation.Extensive experiments on the TAOBAO-MM and KuaiRec datasets demonstrate that GLASS  
outperforms  baselines. In future work, we aim to explore the potential of integrating cross domain long-term behaviors to further enrich the semantic search process.


\bibliographystyle{ACM-Reference-Format}
\bibliography{reference}

\end{document}